\documentclass{svproc}
\usepackage{url}
\usepackage{amsmath}
\usepackage{graphicx}
\usepackage{booktabs}

\begin{document}
\mainmatter              
\title{Robust Coordinated Longitudinal Control of MAV Based on Energy State}
\titlerunning{Longitudinal autopilot}  
%
\author{Chenlong Zhang\inst{1,2} \and Dawei Li\inst{2} \and Haodong Li\inst{1,2}}
\authorrunning{Chenlong Zhang et al.} 
%
\tocauthor{Chenlong Zhang, Dawei Li, and Haodong Li}
\institute{School of Automation Science and Electrical Engineering, Beihang University, Beijing 100191, China\\
\and Institute of Unmanned System, Beihang University, Beijing 100191, China}


\maketitle              

\begin{abstract}
Fixed-wing Miniature Air Vehicle (MAV) is not only coupled with longitudinal motion, but also more susceptible to wind disturbance due to its lighter weight, which brings more challenges to its altitude and airspeed controller design. Therefore, in this paper, an improved longitudinal control strategy based on energy state, is proposed to address the above-mentioned issues. The control strategy utilizes the Linear Extended State Observer (LESO) to observe the energy states and the disturbance of the MAV, and then designs a Multiple-Input Multiple-Output (MIMO) controller based on a more coordinated Total Energy Control (TEC) strategy to control the airspeed and altitude of the MAV. The performance of this control strategy has been successfully verified in a Model-in-the-Loop (MIL) simulation with Simulink, and a comparative test with the classical TEC algorithm is carried out.
\end{abstract}
\section{Introduction}

MAV specifically refers to fixed-wing Unmanned Aerial Vehicles (UAVs) with a wingspan of less than 5 feet\cite{Beard:Randal W.}. The conventional longitudinal control system of MAV adopts the design method of Single-Input Single-Output (SISO), such as the altitude and the airspeed are controlled by the elevator and the throttle, respectively. However, this approach results in uncoordinated longitudinal control and insufficient flight envelope protection\cite{Yayla}.

In order to overcome the inherent shortcomings of the classical SISO controller mentioned above, Lambregts proposed the concept of TEC, which designed a longitudinal controller of MAV from the perspective of controlling the variation and distribution of the total energy of MAV\cite{Lambregts A}. Then Faleiro et al. analyzed the superiority of the MIMO controller based on TEC, and designed an improved TEC controller using the method of eigenstructure assignment\cite{Faleiro LF}. Based on the research of Faleiro and Lambregts, QZ Zhang et al. introduced deviation control into the TEC controller, and realized the zero steady-state error control of the altitude and airspeed\cite{Zhang QZ:An JW:Liu XG,Zhang QZ:An JW}.

Nowadays, the fixed-wing longitudinal controllers in the famous open source projects Ardupilot\footnote{https://ardupilot.org/} and PX4\footnote{https://px4.io/} for UAVs are all designed based on TEC. And many researchers also combine TEC with other control strategies to improve controller performance. Ying-Chih Lai et al. accomplished Optimal Energy Control System (OECS) using Linear-Quadratic-Gaussian (LQG) and gain scheduling control\cite{Lai YC}. Brigido-González et al. proposed a nonlinear description of TEC, and adopted an adaptive strategy to design the TEC controller\cite{Brigido}. Degaspare et al. proposed a more integrated TEC control framework using robust control theory by adding engine related variables to TEC architecture\cite{Giusti Degaspare}. Furthermore, Elgohary et al. completed the hardware-in-the-loop(HIL) simulation of the TEC-based altitude hold controller using Simulink\cite{Elgohary}, and Jimenez et al. completed the real-world test of the TEC controller using a low-cost MAV\cite{Jimenez}.

However, the focus of the above methods does not consider the effect of turbulence, which is a challenge that can not be ignored for MAVs. Therefore, in this paper, we explore the transformation between the energy states and the control actuators, and improve the basic architecture of TEC in conjunction with Linear Active Disturbance Rejection Control (LADRC) for stronger robustness. And the superiority of the designed control scheme, which is subsequently called LADRC-TEC, is demonstrated by comparison with the classical TEC scheme.

The remainder of this paper is structured as follows. Section 2 briefly describes the principles of TEC and LADRC. Section 3 derives energy state differential equations of the MAV, and details the internal structure of the LADRC-TEC. Section 4 completes the simulation experiments through Simulink. Finally, Section 5 summarizes the work of the full paper and indicates the future work.

\section{The Principle of TEC and LADRC}

\subsection{Longitudinal Motion and Coupling Characteristics of MAV}






The longitudinal kinematic equations of the MAV in its body frame\cite{Zhang ML,Wu ST} are:

\begin{equation}\label{1}
    \left\{
    \begin{array}{l}
        \dot{h} = u \sin \theta-v \sin \phi \cos \theta-w \cos \phi \cos \theta \\
        \dot{u} = r v-q w-g \sin \theta + \frac{A_x^b +T}{m} \\
        \dot{w} = q u-p v+g \cos \theta \cos \phi+ \frac{A_z^b}{m} \\
        \dot{\theta} = q \cos \phi-r \sin \phi \\
        \dot{q} = \Gamma_{5} p r-\Gamma_{6}\left(p^{2}-r^{2}\right)+\frac{M_{pitch}}{J_y} 
    \end{array}
    \right.
\end{equation}
where $h$ represents the altitude of the MAV; $[u, v, w]^T$ and $[p, q, r]^T$ are the linear velocities and the angular velocities of the MAV, respectively; $A_x^b$ and $A_z^b$ represent the components of the aerodynamic force under the x-axis and z-axis of the body frame, respectively; $T$ represents the thrust of the propeller; $m$ represents the total weight of the MAV; $\theta$ and $\phi$ represent the pitch angle and the roll angle, respectively.

$A_x^b$ and $A_z^b$ are related to the lift $L$ and the drag $D$ as follows:
\begin{eqnarray*}
    A_x^b = L sin\alpha-D cos\alpha,A_z^b = -L cos\alpha-D sin\alpha
\end{eqnarray*}
where $\alpha$ represents the angle of attack.

Under the condition of stable level flight without wind, there is $\phi=v=p=r=0$. Because $\alpha=tan^{-1}(\frac{w}{u})\approx0$, there is $w \approx 0$, at this time the airspeed $V_a=u$. 

Then Eq.(\ref{1}) can be simplified as:

\begin{equation}\label{2}
    \left\{
    \begin{array}{l}
        \dot{h}\ \ = V_a sin \theta \\
        \dot{V_a} = \frac{T-D}{m}  -g \sin \theta\\
    \end{array}
    \right.
\end{equation}
It can be seen from Eq.(\ref{2}) that for the MAV, the change of pitch angle will lead to the change of MAV's altitude and airspeed; the change of thrust will lead to the change of MAV's airspeed, and subsequently lead to the change of MAV's altitude.

Due to these kinematically coupled characteristics of the MAV, SISO-based controllers will lead to inconsistencies in control, resulting in excessive waste of energy and longer response times.

\subsection{The Principle of TEC}

The TEC decouples the altitude and airspeed control by controlling and distributing the total energy of the MAV. The throttle and the elevator are utilized to control the total energy variation  and the  total energy distribution of the MAV, respectively.

The total energy of the MAV includes the kinetic energy and gravitational potential energy of the MAV:

\begin{equation}\label{3}
E_{T}=E_{D}+E_{S}=\frac{1}{2} m V_a^{2}+m g h
\end{equation}
Then, the total energy rate of MAV is:

\begin{equation}\label{4}
\dot{E}_{T}=m V_a \dot{V_a}+mg \dot{h}
\end{equation}
Considering that the pitch angle of MAV is generally relatively small when flying smoothly, we assume $sin\theta \approx \theta$. Then from Eq.(\ref{2}), it can be deduced that the required thrust during flight is:

\begin{equation}\label{5}
    T=m g\left(\frac{\dot{V_a}}{g}+\theta\right)+D
\end{equation}
From Eq.(\ref{2}), Eq.(\ref{4}) and Eq.(\ref{5}), we can get:

\begin{equation}\label{6}
\dot{E}=\frac{\dot{E}_{T}}{mg V_a}=\frac{\dot{V_a}}{g}+\theta=\frac{T-D}{m g}
\end{equation}
where $\dot{E}$ represents the specific energy rate.

Assuming that the drag of the UAV is constant during flight, then the change of thrust of the MAV is only related to the change of pitch angle and airspeed of MAV. Consequently, the following relationship is formulated:

\begin{equation}\label{7}
\Delta T=m g\left(\frac{\Delta \dot{V_a}}{g}+\Delta \theta\right) \propto \Delta \dot{E}
\end{equation}
where $\Delta T$ represents the change in thrust, $\Delta {V_a}$ represents the change in airspeed, $\Delta \theta$ represents the change in pitch angle, $\Delta \dot{E}$ represents the change in specific energy rate.

From this, we can obtain the control law based on the PI controller, which controls the thrust change by controlling the total energy change of the MAV:	

\begin{equation}\label{8}
T^{d}=k_{E p}\Delta \dot{E}+\int_{t_0}^t k_{E i}\Delta \dot{E} dt+ T^{*}
\end{equation}
where $T^d$ represents the desired thrust after the flight state changes; $T^*$ is the nominal thrust of the MAV in the cruise phase; $k_{E p}$ and $k_{E i}$ represent the proportional and integral gain, respectively.

Analogously, the concept of specific energy distribution rate is introduced for controlling the total energy distribution, which is defined as:

\begin{equation}\label{9}
    \dot{B}=\theta-\frac{\dot{V_a}}{g}
\end{equation}
Similar to the above control law, the transitive relationship between the desired pitch angle $\theta^{d}$ and the change of specific energy distribution rate $\Delta \dot{B}$ is expressed as follows:

\begin{equation}\label{10}
    \theta^{d}=k_{B p}\Delta \dot{B}+\int_{t_0}^t k_{B i}\Delta \dot{B} dt
\end{equation}
where $\theta^d$ represents the desired pitch angle after the flight state changes; $k_{B p}$ and $k_{B i}$ represent the proportional and integral gain, respectively.

In order to avoid introducing unnecessary zeros into the control system, combining Eq.(\ref{8}) and Eq.(\ref{10}), the core algorithm of TEC theory can be simplified as \cite{Faleiro LF}:

\begin{equation}\label{11}
    \left\{\begin{array}{l}
    T^{d}=k_{E p}\dot{E}+ \int_{t_0}^t k_{E i}\Delta \dot{E} dt  + T^{*} \\
    \theta^{d}=k_{B p} \dot{B}+\int_{t_0}^t k_{B i}\Delta \dot{B} dt
    \end{array}\right.
\end{equation}

\subsection{The Principle of LADRC}

\begin{figure}[htbp]
	\centering
	\includegraphics[width=0.7\textwidth]{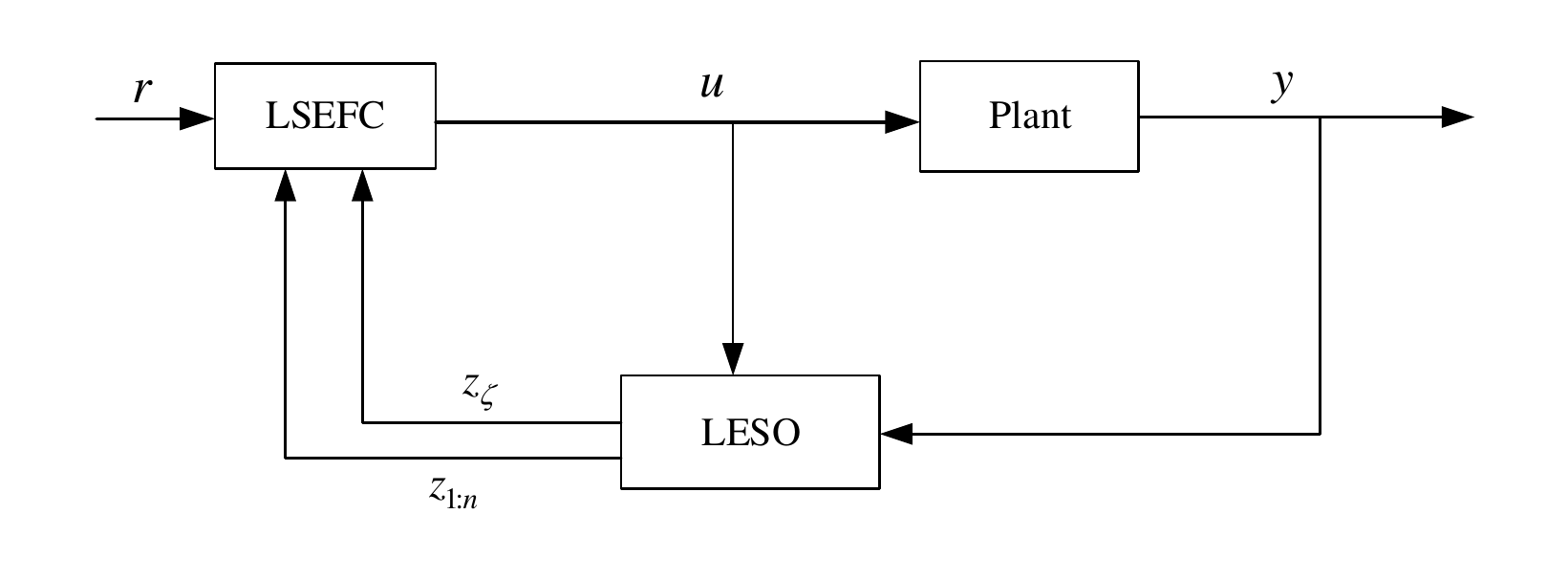}
	\caption{Architecture of LADRC. $z_{1:n}$ and $z_{\zeta}$ represent the original state variables and observed disturbances of the plant, respectively; $y$ and $u$ are the output and input of the plant, respectively; $r$ is the input of the controller.}
	\label{fig:-Architecture of LADRC}
\end{figure}

The LADRC is proposed by ZQ Gao, which mainly includes LESO and linear state error feedback controller (LSEFC)\cite{Gao Zhiqiang}. LESO is utilized to observe the total disturbance of the plant and expand it into a new state variable. LSEFC adopts the feedforward control strategy to compensate the total disturbance of the plant based on the order of the plant. Its control architecture is shown in Fig.\ref{fig:-Architecture of LADRC}.

\noindent Next, the structure of LADRC is further elaborated through a second-order system. Let the controlled object be:

\begin{equation}\label{12}
\ddot{y}=a_1\dot{y}+a_0 y+d+bu
\end{equation}
where $d$ is the disturbance.

By setting $d$ to the extended state, which means that the state variable is $\boldsymbol{x}=\begin{bmatrix} y & \dot{y} & d \end{bmatrix}^T$, Eq.(\ref{12}) can be converted into a continuous extended state space description:

\begin{equation}\label{13}
    \left\{
    \begin{array}{rcl}
        \dot{\boldsymbol{x}} &=& \boldsymbol{A}\boldsymbol{x}+\boldsymbol{B}u+\boldsymbol{E}\dot{d}\\
        y &=& \boldsymbol{C}\boldsymbol{x} \\
    \end{array}
    \right.
\end{equation}
where $\boldsymbol{A}=\begin{bmatrix}0 & 1 & 0\\ a_0 & a_1 & 1 \\ 0 & 0 & 0 \end{bmatrix}$, $\boldsymbol{B}=\begin{bmatrix}0 \\ b \\ 0  \end{bmatrix}$, $\boldsymbol{E}=\begin{bmatrix}0 \\ 0 \\ 1  \end{bmatrix}$, $\boldsymbol{C}=\begin{bmatrix}1 & 0 & 0  \end{bmatrix}$.

Let the state variable of the observer be $\boldsymbol{\hat{x}} $, and the corresponding LESO is:

\begin{equation}\label{14}
LESO:
    \left\{
    \begin{array}{l}
\hat{\dot{\boldsymbol{x}}}  \ = \left(\boldsymbol{A}-\boldsymbol{L}\boldsymbol{C}\right)\boldsymbol{\hat{x}}+\left[\boldsymbol{B},\boldsymbol{L}\right]\boldsymbol{u}_c\\
        \boldsymbol{y}_c = \boldsymbol{\hat{x}} \\
    \end{array}
    \right.
\end{equation}
where $\boldsymbol{u}_c=\begin{bmatrix}u & y \end{bmatrix}^T$ is the input variable of LESO, $\boldsymbol{y}_c=\begin{bmatrix} \hat{y} & \hat{\dot{y}} & \hat{d} \end{bmatrix}^T$ is the output variable of LESO, $\boldsymbol{L}$ is the observer gain matrix to be designed. 

Note that $\dot{d}$ is omitted in Eq.(\ref{14}), because $\dot{d}$ can be estimated by the correction term.

LSEFC can be regarded as a feedforward based Proportional-Integral (PD) controller, which can be expressed as:

\begin{equation}\label{15}
LSEFC:
\begin{cases}
u_{0}=k_{p}\left(r-\hat{y}\right)-k_{d} \hat{\dot{y}} \\ u \ =\left(u_{0}-\hat{d}\right) / {b}
\end{cases} 
\end{equation}

\section{The Design of MAV's Longitudinal Controller}

\begin{figure}[htbp]
	\centering
	\includegraphics[width=1.0\textwidth]{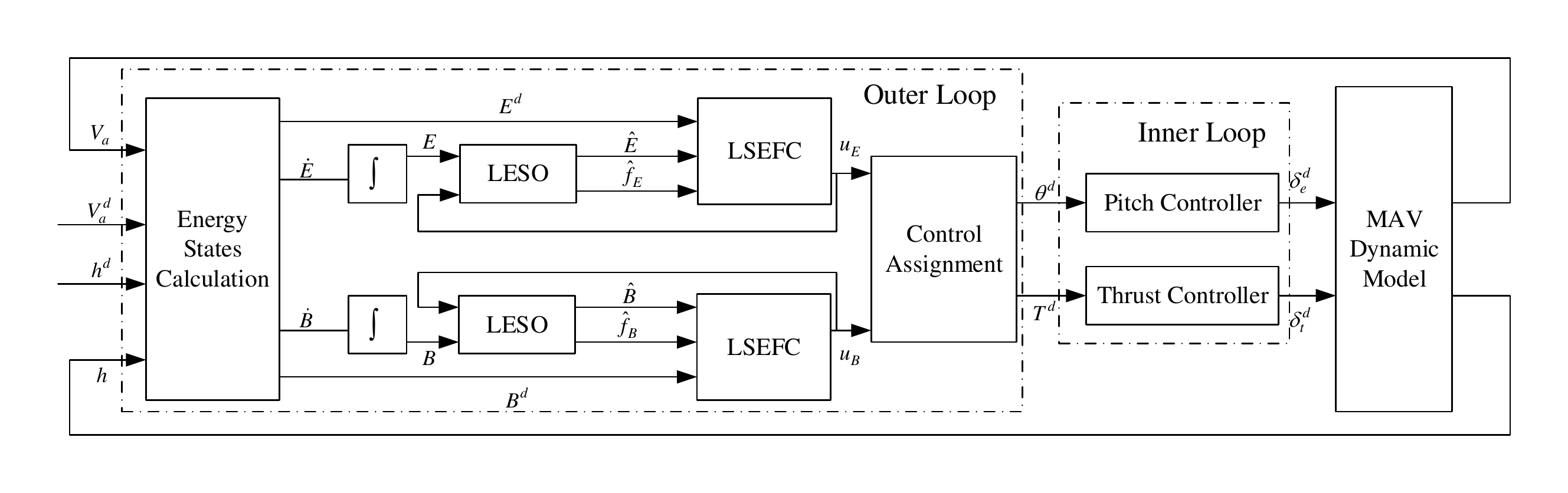}
	\caption{Architecture of LADRC-TEC. $(\ )^d$ represents the desired value. $(\ \hat{} \ )$ represents the estimate of LESO.}
	\label{fig:-Control system structure diagram}
\end{figure}

The longitudinal controller of MAV consists of two parts: the outer loop and the inner loop. In the outer loop, we design the LESO to observe energy states of the MAV and the corresponding disturbance, based on dynamic differential equations of $\dot{E}$ and $\dot{B}$ of the MAV, which reveal the inherent flaws of traditional TEC. And then we utilize a MIMO controller, which has a more coordinated transfer relationship between energy states and control quantities, to calculate the desired pitch angle and thrust. The inner loop is used to pass the desired pitch angle and thrust commands to the elevator and the throttle. The overall control system architecture is shown in Fig.\ref{fig:-Control system structure diagram}.

\subsection{The Design of Inner Loop Controller}

\subsubsection{The Design of Thrust Controller}

Usually, MAV is an electric aircraft, and the corresponding thrust is generated by the rotation of the propeller, so a simplified propeller thrust model can be generated by using Bernoulli's principle\cite{Argyle M E}:

\begin{equation}\label{16}
T =\frac{1}{2}\rho S_{prop} C_{prop}  \left(k_{motor}^2 \delta_t^2-V_a^2 \right) =k_{T1}\delta_t^2-k_{T2}V_a^2
\end{equation}
where $k_{T1}$ and $k_{T2}$ are constant parameters related to the propeller motor.

So a feedback-free thrust controller can be expressed as:

\begin{equation}\label{17}
\delta_{t}^d = \sqrt{\frac{T^d+k_{T2}V_a^2}{k_{T1}}}
\end{equation}
The most significant advantage of using this kind of thrust controller is that we can skip the MAV's thrust control command and directly generate the MAV's throttle control command when designing the outer loop controller.

\subsubsection{The Design of Pitch Controller}

To simplify the controller design process, the pitch controller is implemented with a PID controller:

\begin{equation}\label{18}
\delta_{e}^d=k_{p, \theta}\left(\theta^{d}-\theta\right)+k_{i, \theta} \int_{t_{0}}^{t}\left(\theta^{d}-\theta\right) dt -k_{d, \theta} q
\end{equation}

\subsection{The Design of Outer Loop Controller}

In the condition of stable flight with no wind and no sideslip, there is $V_a=\sqrt{u^2+w^2}$, so

\begin{equation}\label{19}
    \begin{aligned}
    \dot{V_a}  = \frac{u \dot{u} + w \dot{w}}{V_a}
    =\dot{u}cos \alpha + \dot{w}sin\alpha
    \end{aligned}
\end{equation}
Substituting into Eq.(\ref{1}), we have

\begin{equation}\label{20}
    \begin{aligned}
    \dot{V_a}  =& cos \alpha \left\{ r v-q w-g \sin \theta + \frac{A_x^b +T}{m} \right\}+
    \\& sin \alpha \left\{ q u-p v+g \cos \theta \cos \phi+ \frac{A_z^b}{m} \right\}  \\
    =& -g sin(\theta - \alpha)-g sin\alpha cos \theta (1- cos\phi)+ \frac{T cos\alpha -D}{m} \\
    =& -g sin(\theta - \alpha)+\frac{T -D}{m}+d_{V}
    \end{aligned}
\end{equation}
where $d_V=-g sin\alpha cos \theta (1- cos\phi)+ \frac{T}{m}(cos\alpha-1)$ is regarded as the disturbance of the system. In the condition of level straight flight, there is $d_V \approx 0$.

Usually the drag can be expressed as $D=\frac{1}{2}\rho V_a^2 S C_D$, so Eq.(\ref{20}) can be linearized as:

\begin{equation}\label{21}
    \begin{aligned}
    \dot{\widetilde{V_a}}  
    =& -g cos(\theta^{*} - \alpha^{*})\widetilde{\theta}+ \frac{2k_{T1}\delta_t^{*}}{m}\widetilde{\delta_t} +\frac{\rho S C_D V_a^{*}-2k_{T2}V_a^{*}}{m}\widetilde{V_a}\\
    & +d_{V}+R_n \\
    =& a_{1}\widetilde{\theta}+a_{2}\widetilde{\delta_t} + a_{3}\widetilde{V_a}+d_{V}+R_n
    \end{aligned}
\end{equation}
where $\dot{\widetilde{V_a}} =\dot{V_a}-\dot{V_a^*}=\dot{V_a}$, $\widetilde{\theta}=\theta-\theta^*$, $\widetilde{\delta_t}=\delta_t-\delta_t^*$, $\widetilde{V_a}=V_a-V_a^*$; $a_{1}, a_{2}$ and $a_{3}$ are constants that can be calculated from the aerodynamic parameters of the MAV; $R_n$ is the remainder of the Taylor polynomial. $V_a^*,\theta^*$ and $\delta_t^*$ are the nominal values at cruise.

From Eq.(\ref{1}), we get:

\begin{equation}\label{22}
    \begin{aligned}
    \dot{h} &= u \sin \theta-v \sin \phi \cos \theta-w \cos \phi \cos \theta \\
    &=V_a \theta + (u sin\theta-V_a \theta)-v \sin \phi \cos \theta-w \cos \phi \cos \theta \\
    & = V_a \theta +d_h
    \end{aligned}
\end{equation}
where $d_h$ is regarded as the disturbance of the system, and $d_h \approx 0$ in the condition of level straight flight.

Therefore
\begin{equation}\label{23}
    \begin{aligned}
    \dot{E} &= \frac{\dot{V_a}}{g}+ \frac{\dot{h}}{V_a}\\
    =&\frac{a_{1}}{g}(\theta-\theta^*)+\frac{a_{2}}{g}(\delta_t-\delta_t^*) + \frac{a_{3}}{g}(V_a-V_a^*)+\frac{d_{V}+R_n}{g}+\theta+\frac{d_h}{V_a}\\
    =&a_{E1}\theta+a_{E2}\delta_t+f_E\\
    =&b_E u_E+f_E
    \end{aligned}
\end{equation}
where $u_E$ is the composite input about $\theta$ and $\delta_t$, $f_E$ is the total disturbance with respect to $\dot{E}$ including external disturbance and internal disturbance.

For the first-order system represented by Eq.(\ref{23}), LESO can be designed as:
\begin{equation}\label{24}
    \left\{
    \begin{array}{l}
        e_E = E- \hat{E}\\
        \hat{\dot{E}} \ =\hat{f_E}+b_Eu_E+L_1e_E\\
        \hat{\dot{f_E}}=L_2e_E\\
    \end{array}
    \right.
\end{equation}
where $L_1$ and $L_2$ are the observer gains to be designed.

After $E$ and $f_E$ are both observed by LESO, LSEFC can be designed as:
\begin{equation}\label{25}
    \left\{
    \begin{aligned}
        u_{E0} &= k_E(E^d- \hat{E})\\
        u_E & = u_{E0} - \frac{\hat{f_E}}{b_E}
    \end{aligned}
    \right.
\end{equation}
where $k_E$ and $b_E$ are adjustable controller gains, $E^d=\int_{t_0}^{t}\left(\frac{\dot{V_{a}^d}}{g}+ \frac{ \dot{h^d}}{V_a}\right)dt$ is the expected specific energy.

The desired airspeed change rate $\dot{V_{a}^d}$ and the desired altitude change rate $\dot{h^d}$ can be calculated by the following equations:
\begin{equation}\label{26}
    \left\{
    \begin{aligned}
      \dot{V_{a}^d} &= k_V(V_{a}^d- V_a)\\
        \dot{h^d} & = k_h(h^d-h)
    \end{aligned}
    \right.
\end{equation}
where $k_V$ and $k_h$ are gain factors.

Similarly, for the specific energy distribution rate $\dot{B}$, there is:
\begin{equation}\label{27}
    \begin{aligned}
    \dot{B} &= -\frac{\dot{V_a}}{g}+ \frac{\dot{h}}{V_a}\\
    &=a_{B1}\theta+a_{B2}\delta_t+f_B\\
    &=b_B u_B+f_B
    \end{aligned}
\end{equation}
where $u_B$ is the composite input about $\theta$ and $\delta_t$, $f_B$ is the total disturbance with respect to $\dot{B}$ including external disturbance and internal disturbance.

The design process of specific energy distribution rate's LESO and LSEFC is similar to that of $\dot{E}$, and will not be repeated here.

From Eq.(\ref{23}) and Eq.(\ref{27}), it can be seen that:

\begin{equation}\label{28}
    \begin{bmatrix}
    b_E u_E\\
    b_B u_B
    \end{bmatrix}
    =
    \begin{bmatrix}
    a_{E1} & a_{E2} \\
    a_{B1} & a_{B2}
    \end{bmatrix}
    \begin{bmatrix}
    \delta_t\\
    \theta
    \end{bmatrix}
    =\boldsymbol{A}
    \begin{bmatrix}
    \delta_t\\
    \theta
    \end{bmatrix}
\end{equation}
where $\boldsymbol{A}$ plays a role similar to the control allocation matrix.

Therefore

\begin{equation}\label{29}
    \begin{bmatrix}
    \delta_{t}^d\\
    \theta^d
    \end{bmatrix}
    =\boldsymbol{A}^{-1}
    \begin{bmatrix}
    b_E u_E\\
    b_B u_B
    \end{bmatrix}
\end{equation}
which means that there is no one-to-one correspondence between $\dot{E}$, $\dot{B}$ and $\delta_{t}^d$, $\theta^d$. 

In other words, the total energy variation and distribution of the MAV can not be simply controlled by the throttle and the pitch angle, respectively, which is also the defect of the traditional TEC theory. Using the control allocation scheme shown in Eq.(\ref{29}) will obtain a more coordinated control effect, which will allow the MAV to save more control energy and achieve faster response.

\section{Simulation Verification and Result Analysis}

Through MIL simulation in Simulink, LADRC-TEC will be verified by comparison with the classical total energy control strategy expressed by Eq.(\ref{11}), which is subsequently referred to as simply TEC. The purpose of the simulation experiment is to verify the performance of the designed controller from multiple perspectives, including dynamic response, wind resistance and energy consumption. The simulation will build a six-degree-of-freedom nonlinear full-scale equation of the MAV in Simulink, and the lateral controller will be implemented using a simple PD controller. The aerodynamic parameters of the MAV used are from \cite{Beard:Randal W.}, which is the Aerosonde UAV's. 

And using the aerodynamic parameters of the Aerosonde UAV, we can calculate $\boldsymbol{A}=\begin{bmatrix}5.7639 & -0.0003 \\ -5.7639 & 2.0003 \end{bmatrix}$ in Eq.(\ref{28}). This indicates that for the Aerosonde UAV, $\dot{E}$ is basically affected by throttle, while $\dot{B}$ is affected by both throttle and pitch angle.

\subsection {Simulation Analysis of Step Response}

\begin{figure}[htbp]
	\includegraphics[width=1\textwidth]{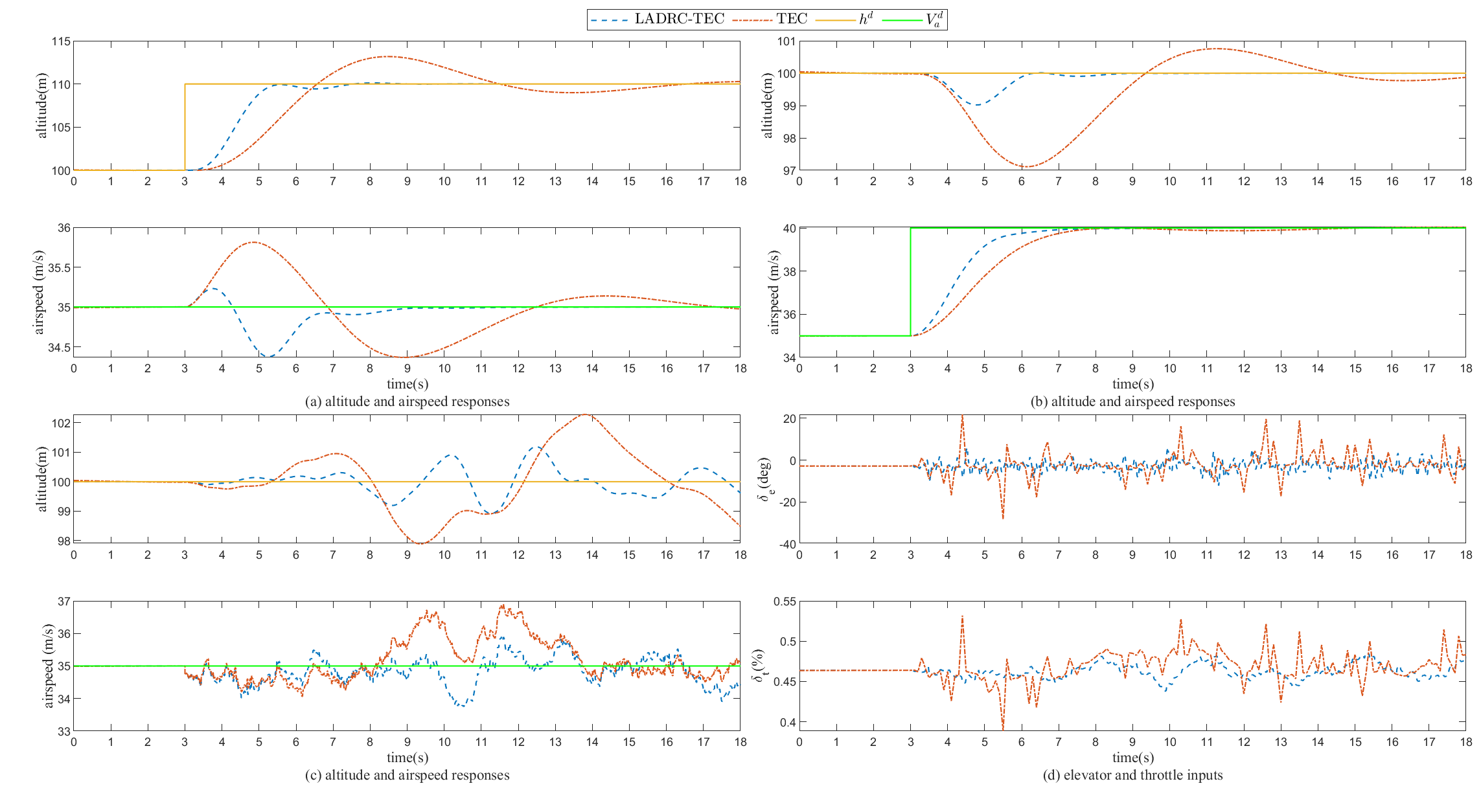}
	\caption{The dynamic response of MAV}
	\label{fig:-Dynamic response of MAV}
\end{figure}

We assume that the airspeed of the MAV is 35m/s, and the altitude of the MAV is 100m during cruise. The altitude step response is to receive the command to climb 10m, and the dynamic response of the MAV is demonstrated in Fig.\ref{fig:-Dynamic response of MAV}a. The airspeed step response is to receive the command to accelerate 5m/s, and the dynamic response of the MAV is demonstrated in Fig.\ref{fig:-Dynamic response of MAV}b. It can be seen that whether it is altitude response or airspeed response, the settling time of LADRC-TEC is shorter than that of TEC, and the overshoot of LADRC-TEC is also smaller than that of TEC, which reflects that LADRC-TEC has a more coordinated control effect.

\subsection {Simulation Analysis of Wind Disturbance Response}

A good model of the turbulent wind field can be obtained by passing the white noise through the shaping filter given by the von Karmen turbulence spectrum. The Dryden transfer function gives an approximation of the von Karmen longitudinal turbulence model as follows\cite{T R Beal}:

\begin{equation}\label{30}
    \left\{
    \begin{aligned}
    &G_{u}(s)=\sigma_{u} \sqrt{\frac{2 V_{a}}{L_{u}}} \frac{1}{s+\frac{V_{a}}{L_{u}}} \\
    &G_{w}(s)=\sigma_{w} \sqrt{\frac{3 V_{a}}{L_{w}}} \frac{\left(s+\frac{V_{a}}{\sqrt{3} L_{w}}\right)}{\left(s+\frac{V_{a}}{L_{w}}\right)^{2}}
    \end{aligned}
    \right.
\end{equation}
where $\sigma_{u}$ and $\sigma_{w}$ represent the turbulence intensity along the x-axis and z-axis of the body frame, respectively; and $L_u$ and $L_w$ are the turbulence scales. These parameters are used to describe different turbulence intensities at different altitudes.

\begin{figure}[htbp]
	\centering
	\includegraphics[width=0.5\textwidth]{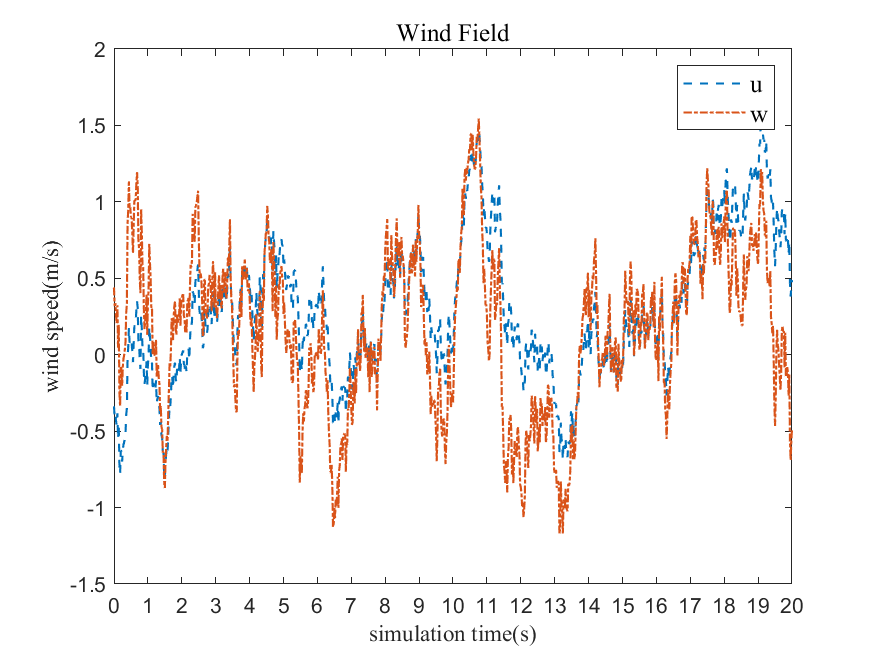}
	\caption{The Velocity of turbulent wind}
	\label{fig:-Turbulent wind speed change diagram}
\end{figure}

To simulate mild turbulence at low altitudes, the above parameters were chosen as $\sigma_{u}=1.06$, $\sigma_{w}=0.7$, $L_u=200$, $L_w=50$. The generated longitudinal turbulent wind field is demonstrated in Fig.\ref{fig:-Turbulent wind speed change diagram}, which indicates that the maximum velocity of turbulent wind does not exceed 2m/s.

In order to evaluate the anti-wind disturbance capability of LADRC-TEC, we set the MAV to cruise steadily in the simulation environment, but suddenly encountered atmospheric turbulence in the 3rd second. The altitude and airspeed responses of the MAV are demonstrated in Fig.\ref{fig:-Dynamic response of MAV}c. The elevator and throttle inputs of the MAV are demonstrated in Fig.\ref{fig:-Dynamic response of MAV}d. Obviously, the altitude and airspeed oscillations and the changes of the control inputs of the MAV using LADRC-TEC strategy are much smaller than those of the MAV using TEC strategy.

If the quantitative analysis method is used, the standard deviation of the MAV's altitude, airspeed, elevator input and throttle input under the nominal cruise condition is calculated. And the results are shown in Table 1, which further validates that LADRC-TEC is more robust.

\begin{table}[htbp]
\centering
\caption{standard deviation of wind disturbance response}
\begin{tabular}{ccccc}
\toprule
 & $h(m)$ & $V_a(m/s)$ & $\delta_e(deg)$ & $\delta_t(\%)$\\
\midrule
LADRC-TEC  &  0.4651& 0.4524& 2.6153 & 0.0088\\
TEC  &   1.0833& 0.6733 & 12.4518 & 0.1208 \\
\bottomrule
\end{tabular}
\end{table}

\section {Conclusion}

In this paper, we derived dynamic differential equations of energy states in TEC theory, which show that the coupled control relationship between the energy states and the longitudinal control actuators. Then based on LESO, we further observed the energy states of the MAV and the corresponding disturbances, which are converted into the desired pitch angle and throttle through the MIMO controller. Through Simulink simulation, we demonstrated that this control strategy has better performance compared with the original TEC control strategy.

However, the performance of the designed controller has only been subjected to preliminary simulation verification, and its feasibility has not been verified in engineering practice.

%
%

\end{document}